\documentclass[aps,prev,preprintnumbers,floatset,nofootinbib,twocolumn]{revtex4-1}
\pdfoutput=1

\usepackage[english]{babel}
\usepackage{bm}
\usepackage{times}
\usepackage{ulem}
\usepackage{hyperref}
\hypersetup{
    colorlinks=true,
    linkcolor=blue,
    filecolor=magenta,      
    urlcolor=blue,
    citecolor=blue,
    pdftitle={Sharelatex Example},
    pdfpagemode=FullScreen
    }
\usepackage{listings}
\usepackage{slashed}
\usepackage{color}
\usepackage{appendix}
\usepackage{mathtools}
\usepackage{epsfig}
\usepackage{dcolumn}
\usepackage{textcomp}
\usepackage{appendix} 
\usepackage{multirow}
\usepackage{graphicx}
\usepackage{soul}
\usepackage{subfigure}
\usepackage{xcolor}
 \usepackage{ulem}
\usepackage{enumerate}
\usepackage[capitalize]{cleveref}

\newcommand{\eq}[1]{Eq.~(\ref{#1})} 
\newcommand{\tab}[1]{Table~\ref{#1}} 
\newcommand{\zp}{Z^{\prime}}

\begin{document}

\title{Flavor changing interactions confronted with meson mixing and hadron colliders}

\author{A. E. C\'arcamo Hern\'andez$^{1,2,3}$}
\email{antonio.carcamo@usm.cl}
\author{L. Duarte$^{4}$}
\email{laura.duarte.ara@gmail.com}
\author{A. S. de Jesus$^{4,5}$}
\email{alvarosndj@gmail.com}
\author{S. Kovalenko$^{2,3,6}$}
\email{sergey.kovalenko@unab.cl}
\author{F. S. Queiroz$^{3,4,5}$}
\email{farinaldo.queiroz@ufrn.br}
\author{C. Siqueira$^{7}$}
\email{csiqueira@ifsc.usp.br}
\author{Y.M. Oviedo-Torres$^8$}
\email{ymot@estudantes.ufpb.br}
\author{Y. Villamizar$^{4,5}$}
\email{Corresponding author: yoxarasv@gmail.com}

\affiliation{$^1$Universidad T\'ecnica Federico Santa Mar\'ia, Casilla 110-V, Valparaiso, Chile.}
\affiliation{$^2$Centro Cient\'\i fico 	Tecnol\'ogico de Valpara\'\i so-CCTVal,	Universidad T\'ecnica Federico Santa Mar\'\i a, Casilla 110-V, Valpara\'\i so, Chile}
\affiliation{$^3$Millennium Institute for Subatomic Physics at the High-Energy Frontier (SAPHIR) of ANID, Fern\'andez Concha 700, Santiago, Chile}
\affiliation{$^4$International Institute of Physics, Universidade Federal do Rio Grande do Norte,
Campus Universitario, Lagoa Nova, Natal-RN 59078-970, Brazil}
\affiliation{$^{5}$Departamento de F\'isica, Universidade Federal do Rio Grande do Norte, 59078-970, Natal, RN, Brasil}
\affiliation{$^6$Departamento de Ciencias F\'isicas, 
Universidad Andres Bello, Sazi\'e 2212, Piso 7, Santiago, Chile} 
\affiliation{$^7$Instituto de F\'isica de S\~ao Carlos, Universidade de S\~ao Paulo, Av. Trabalhador S\~ao-carlense 400, S\~ao Carlos-SP, 13566-590, Brasil.}
\affiliation{$^8$Departamento de Fisica, Universidade Federal da Paraiba, Caixa Postal 5008, 58051-970, Joao Pessoa, PB, Brazil}

\begin{abstract}
We have witnessed some flavor anomalies appeared in the past years, and explanations based on extended gauge sectors are among the most popular solutions. These beyond the Standard Model (SM) theories often assume flavor-changing interactions mediated by new vector bosons. Still, at the same time, they could yield deviations from the SM in the  $K^{0}-\bar{K}^{0}$, $D^{0}-\bar{D}^{0}$, $B^0_d-\bar{B^0}_d$ and $B^0_s-\bar{B^0}_s$ meson systems. Using up-to-date data on the mass difference of these meson systems, we derive lower mass bounds on vector mediators for two different parametrizations of the quark mixing matrices. Focusing on a well-motivated model based on the fundamental representation of the weak SU(3) gauge group, we put our findings into perspective with current and future hadron colliders to conclude that meson mass systems can give rise to bounds much more stringent than those from high-energy colliders and that recent new physics interpretations of the $b\rightarrow s$ and $R(D^{\ast})$ anomalies are disfavored.
\end{abstract}

\maketitle

\section{Introduction}

Since flavor-changing neutral current (FCNC) processes are forbidden at tree-level in the Standard Model (SM), they are very sensitive to new physics. For this reason,  meson-antimeson mixing that belongs to the class of flavor-changing neutral current (FCNC) processes are great laboratories for flavor-changing interactions. Meson systems are key to our understanding of the fundamental interactions and continuously give rise to important results such as the recent measurement of mixing and CP violation in neutral charm mesons collected by the LHCb experiment \cite{LHCb:2021ykz}.

FCNCs have historically been important to the development of SM. From the considerations of FCNC, the charm quark was predicted to accommodate the data that ruled out larger FCNC effects \cite{Politzer:1973fx}. Analyzing the neutral kaon meson system,  the value of charm mass was estimated \cite{Glashow:1970gm}. Charged Kaon decays revealed that weak interactions do not conserve Parity and Charge operators. Moreover,  the $K_L$ decay into pions has shown that CP is not preserved \cite{Gaillard:1974hs}. The SM with two fermion generations could not reproduce this decay because CP-violating interactions of quarks necessarily involve complex couplings. Those complex couplings, if introduced in a $2 \times 2$ mixing matrix, are eliminated after rotation, leaving, in the end, a real $2 \times 2$ Cabibbo matrix. Kobayashi and Maskawa concluded in 1973 that such complex terms would survive in the quark mixing matrix if there were at least three generations. This fact was ignored for quite some time until the discovery of the bottom quark in 1977 by Ledermann \cite{Herb:1977ek}, which later hinted at the existence of the top quark \cite{CDF:1995wbb,D0:1995jca}. 
\begin{figure}[!h]
    \centering
    \includegraphics[scale=0.2]{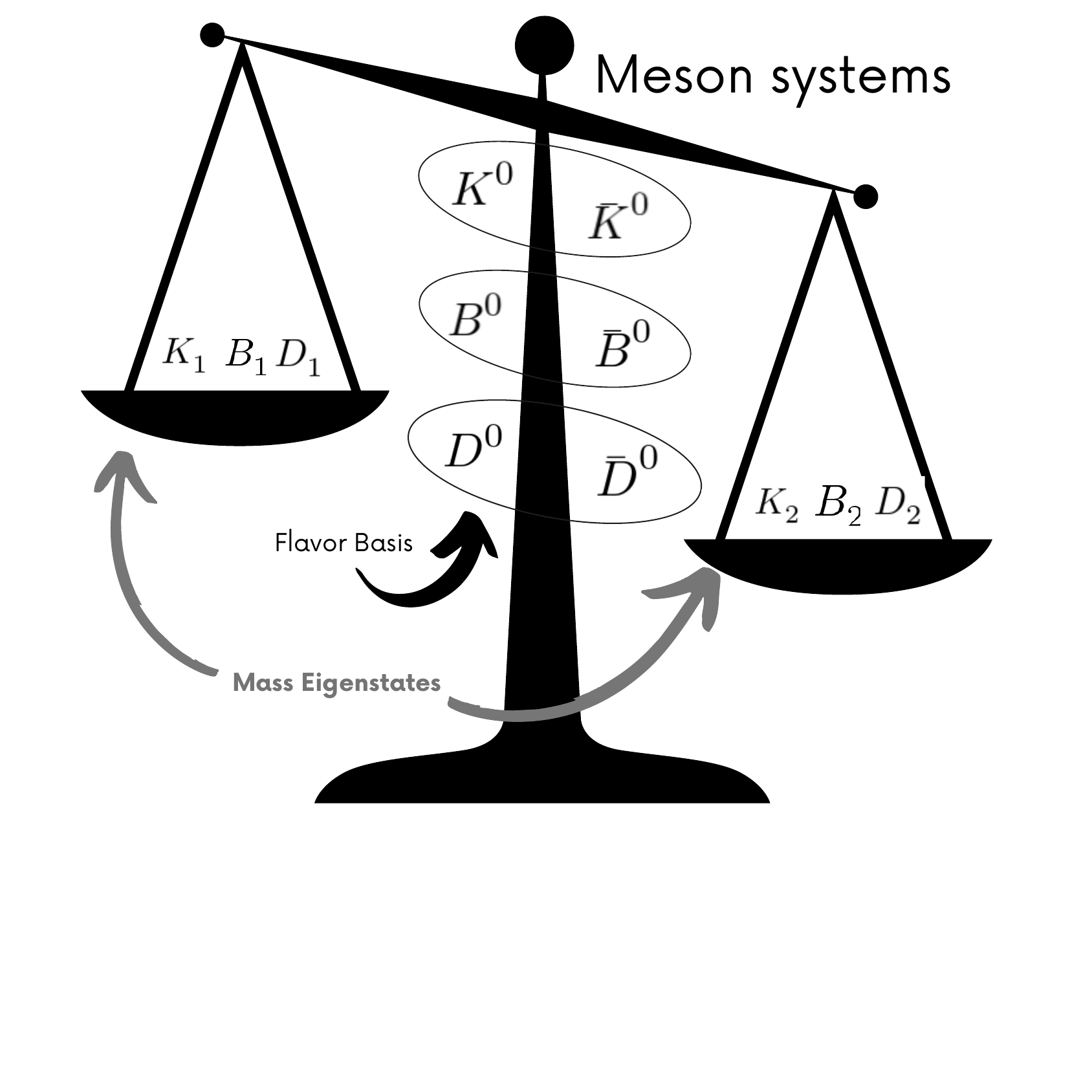}
    \caption{ An illustration of how flavor eigenstates of mesons lead to mass eigenstates of mesons with different masses. The mass difference in such systems of mixed mesons is at the core of our study.
    %
    }
    \label{figmesons}
\end{figure}
Therefore, it is clear that mesons have played a crucial role in our understanding of fundamental interactions. As they are made up of one quark and one antimatter quark. The antimatter state of a given meson is also comprised of a quark and one antimatter quark. For instance, the $D_0$ meson consists of a charm quark and an up antiquark, whereas its antiparticle, the $\bar{D}_0$, is made of a charm antiquark and an up quark. In the quantum physics world, the meson $D_0$ particle can be itself and its antiparticle at once, leading to a quantum superposition of states, say $D_1$ and $D_2$, each with their own mass and their decay width $\Gamma_1$ and $\Gamma_2$ (See Fig.~\ref{figmesons}). This superposition allows a continuous oscillation between the $D_0$ particle and its antiparticle. In other words, the Hamiltonian is not diagonal in the flavor basis, and thus flavor changing interactions are present. The mass difference, $m_{D1}-m_{D2}$, determines the frequency of oscillations, which is measured \cite{BELLE:2007dgh,BaBar:2007kib,HFLAV:2019otj}, and reported in terms of the dimensionless parameter $x= (m_{D1}-m_{D2})/\Gamma $, where $\Gamma$ is the average width, $(\Gamma_1+\Gamma_2)/2$.

Such an oscillation pattern is present in four well-known meson systems, namely $K^{0}-\bar{K}^{0}$, $D^{0}-\bar{D}^{0}$, $B^0_d-\bar{B^0}_d$ and $B^0_s-\bar{B^0}_s$. The SM FCNC occurs at a one-loop level via a W boson exchange in a box-diagram, involving the Cabibbo–Kobayashi–Maskawa (CKM) matrix; precisely for that reason, any new physics-inducing flavor-changing interactions are tightly constrained by the mesons systems aforementioned.  The relevant quantity for our reasoning is the mass difference between these mesons, where an excellent agreement between theory and measurements is found. In other words, one can use precise measurements on the mass difference of these mesons to constrain any new physics contribution to the mass differences. These mesons are comprised of different quark flavors and, consequently, are sensitive to different entries of the CKM matrix, Therefore, the new physics reach for meson mixing systems relies on the parametrization used for the quark mixing matrices. In summary, a robust assessment of the new physics potential of flavor probes requires control over the systematic errors \cite{Buras:2000dm, Buras:2003jf, D'Ambrosio:2002ex}. 

In our work, we focus on the FCNC effects stemming from neutral vector bosons. A wealth of Abelian and non-Abelian extended gauge symmetries predict the existence of extra neutral gauge bosons \cite{Buras:2014yna}. One can parametrize these new physics contributions in terms of gauge couplings and the mediator mass \cite{Aebischer:2019blw}, but we will concentrate our phenomenology on vector bosons arising from $SU(3)_c \otimes SU(3)_L \otimes U(1)_N$ gauge group, shortly referred to as 3-3-1 models \cite{Singer:1980sw, Pisano:1991ee, Foot:1992rh, Montero:1992jk, Dias:2009au} because models based on this gauge symmetry have been considered as a plausible explanation to the  $b\rightarrow s$ and $R(D^{\ast})$ anomalies \cite{Buras:2013dea, Gauld:2013qja, Descotes-Genon:2017ptp, Wei:2017ago, NguyenTuan:2020xls, Addazi:2022frt}\footnote{See for other flavor studies \cite{Cabarcas:2013jba, Anderson:2005ab}}. FCNC studies in the context of 3-3-1 models have been carried out in the past \cite{Rodriguez:2004mw, Palcu:2007um, Promberger:2007py, Dong:2008sw, Dong:2008sw, Dong:2010zu, Dong:2011vb, Cogollo:2012ek, Giang:2012vs, Hue:2013uw, Machado:2013jca, Hue:2013uw, CarcamoHernandez:2013krw, Vien:2014ica, Vien:2014pta, Hue:2015fbb, Hue:2015fbb, CarcamoHernandez:2015rmj, Vien:2015wca, Machado:2016jzb, Fonseca:2016tbn, Huong:2019vej, Hong:2020qxc, NguyenTuan:2020xls, Oliveira:2022vjo}, but our work differs from previous studies for the following reasons: 
\begin{enumerate}[(i)]
    \item We take into account the four relevant meson systems, including updated measurements;
    \item We consider two different parametrizations to assess the impact of systematic errors; 
    \item As the SM prediction agrees well with the data, we enforce the new physics contribution to be within the reported experimental error bar;
    \item We put our results into perspective with future hadron colliders;
    \item We investigate whether recent proposals based on the 3-3-1 symmetry are consistent with meson mixings and collider bounds.
\end{enumerate}

Our goal is to find lower mass bounds on the vector mediator, a $\zp$, which mediates flavor-changing interactions. Consequently, our findings are relevant to 3-3-1 constructions that feature a similar neutral current with SM quarks \cite{deS.Pires:2007gi,Mizukoshi:2010ky,Alvares:2012qv,Queiroz:2013lca,Kelso:2013nwa,Profumo:2013sca,Kelso:2014qka,Dong:2014esa,Dong:2014wsa,Cogollo:2014jia,Martinez:2014ova,Martinez:2014ova,Martinez:2014rea,daSilva:2014qba,Dong:2015rka,Martinez:2015wrp,Huong:2016ybt,Pires:2016vek}.

Our work is structured as follows: in section {\bf II}, we revise the key ingredients of the 3-3-1 model under study; in section {\bf III}, we derive the 3-3-1 contribution to the mass difference of these mesons; in section {\bf IV}, we discuss the current and future hadron collider bounds; in section {\bf V}, we draw our conclusions.

\section{THE MODEL}

Our FCNC investigation is dedicated to models which are based on the $SU(3)_c \otimes SU(3)_L \otimes U(1)_N$ symmetry, which promotes the SM $SU(2)_L$ gauge group to a $SU(3)_L$ one. There are several ways to arrange fermions in a $SU(3)_L$ triplet, and these multiple possibilities give rise to different 3-3-1 models   \cite{Pisano:1992bxx, Foot:1992rh, Foot:1994ym, Hoang:1996gi, Hoang:1995vq, Ponce:2002sg, Dong:2010zu, Dong:2011vb}. In this work, we will focus on two of the most popular models based on the 3-3-1 symmetry, namely the {\it 3-3-1 model with right-handed neutrinos} (RHN) and the {\it 3-3-1 model with heavy neutral fermion} (LHN). These two particular versions of the 3-3-1 symmetry can accommodate dark matter and neutrino masses, which are the most convincing evidence for physics beyond the SM. We will focus on the 3-3-1 model with right-handed neutrinos, but we emphasize that those two models feature the same neutral current involving the $\zp$ gauge boson and SM quarks. Therefore, our conclusions are valid for both models. That said, under the $SU(3)_c \otimes SU(3)_L \otimes U(1)_N$ gauge group the lepton sector is arranged as,

\begin{equation}
f^a_L=\left(
\begin{array}{c}
\nu^a_l\\
e^a_l\\
(\nu^c_R)^a
\end{array}\right) \sim (1,3,-1/3), \,\, e_R^a \sim (1,1,-1), 
\end{equation} where $a=1,\, 2,\, 3$,  indicate the three fermion generations.

Regarding the hadronic sector, gauge anomaly cancellation requires that the quark generations 
transform differently under $SU(3)_L$ group. The most simple way to accomplish that without invoking several exotic new fermions is by assuming that the first generation transforms as triplets under $SU(3)_L$, whereas the second and third ones as anti-triplets as follows,
\begin{equation}\begin{array}{l}
Q_{3L} = \left(\begin{array}{c}
u_3\\
d_3\\
u_{3}^\prime\end{array}\right)_L \sim (3, 3, 1/3),\\ \\
u_{3R} \sim (3, 1, 2/3),\
d_{3R} \sim (3, 1, -1/3),\ u^\prime_{3R} \sim (3, 1, 2/3),\\ \\
Q_{iL} = \left(\begin{array}{c}
d_i\\
- u_i\\
d_i^\prime\end{array}\right)_L \sim (3, \bar 3, 0), \\ \\
u_{iR} \sim (3, 1, 2/3),\ d_{iR} \sim (3, 1, -1/3),
d_{iR}^\prime \sim (3, 1, -1/3),\end{array} \end{equation}
where $i = 1, \, 2$, with $q^\prime$ being heavy exotic quarks with electric charges $Q(u^\prime_{3})=2/3$ and $Q(d^\prime_{1, \, 2})=-1/3$.

We highlight that in the 3-3-1 LHN, a new heavy neutral lepton $N^{a}_{L}$ replaces the left-handed neutrino in the lepton triplet. In addition, a right-handed neutral fermion $N^{a}_{R} \sim (1,1,0)$ is introduced, which transforms as a singlet under ${SU(3)}_{L}$. The quark sector remains the same though. Hence, as we stressed before, our reasoning for flavor-changing interactions involving quarks applies to both 3-3-1 models.

Fermion masses are generated through the spontaneous symmetry-breaking mechanism governed by three scalar triplets. From a top-down approach, the scalar triplet $\chi$ acquires a vacuum expectation value ($vev$) in the scale of the TeVs with,

\begin{equation}
\langle \chi \rangle = \left(\begin{array}{c}
0\\
0\\
v_\chi\end{array}\right),
\end{equation} breaking $SU(3)_L \otimes U(1)_N$ down to $SU(2)_L \otimes U(1)_Y$, thus generating masses for the additional gauge bosons and new fermions, namely the exotic quarks via the Yukawa Lagrangian,
\begin{equation}{\cal L}^{\chi}_{yuk}= \lambda_1 \bar Q_{1L} u^{\prime}_{1R} \chi
+ \lambda_{2ij} \bar Q_{iL} d^{\prime}_{jR} \chi^{*} + H.c., \end{equation}
where $\chi \sim (1, 3, -1/3)$. 

Then the $SU(2)\otimes U(1)_Y$ breaks into electromagnetism when two scalar triplets $\rho, \eta$ get a $vev$ as follows,

\begin{equation}
\langle \rho \rangle = \left(\begin{array}{c}
0\\
v_{\rho}\\
0\end{array}\right), \
\langle \eta \rangle = \left(\begin{array}{c}
v_{\eta}\\
0\\
0\end{array}\right),
\end{equation} yielding masses for the SM quarks and charged lepton masses through,

\begin{eqnarray}
{\cal L}_{Yuk} && = \lambda_{1a} \bar Q_{1L} d_{aR} \rho
+ \lambda_{2ia} \bar Q_{iL} u_{aR} \rho^{*} + G_{ab}
\bar f^a_L(f^b_L)^c \rho^{*} \nonumber\\
&&+ G^{'}_{ab} \bar f_L^a e_R^b \rho + \lambda_{3a} \bar Q_{1L} u_{aR} \eta +
\lambda_{4ia} \bar Q_{iL} d_{aR} \eta^{*} + H.c.. \nonumber\\
\label{eq:yukawa2}
\end{eqnarray}

Notice that the scalar triplets transform as $\rho \sim (1, 3, 2/3)$ and $\eta \sim (1, 3, -1/3)$. Furthermore, the third term in Eq. (\ref{eq:yukawa2}) gives rise to two mass degenerate neutrinos and a massless one. It is well-known that this neutrino mass pattern cannot reproduce the three mass differences observed in the neutrino oscillation data \cite{Kopp:2013vaa, Gonzalez-Garcia:2015qrr, Bergstrom:2015rba}. However, one can nicely solve this problem by adding a scalar sextet and realizing a type II seesaw mechanism, or adding three right-handed Majorana neutrinos to incorporate an inverse or linear seesaw \cite{Dias:2012xp, Boucenna:2015zwa}. We emphasize that either way neutrino masses are generated, our reasoning concerning FCNC is left unchanged. 

Besides the usual bilinear and quartic terms in the scalar potential, these scalars give rise to the term $-\frac{f}{\sqrt{2}}\epsilon^{ijk}\eta_i \rho_j \chi_k$, where $f$ is in principle a free parameter which has energy dimension. The main energy scale in our work is the energy scale at which the 3-3-1 symmetry is broken down to the SM one. Hence, it is natural to assume that $f \sim v_{\chi}$. 

We highlight this fact, because it is often questioned the importance of FCNC mediated by scalar fields in 3-3-1 constructions. However, if $f\sim v_{\chi}$ the new scalars in the model are heavier than the $Z^\prime$, and consequently the FCNC effects induced by them are relatively smaller than those rising the $Z^\prime$. For concreteness, taking $f=v_{\chi}$, the scalars that induce FCNC have masses larger than $v_{\chi}$. In contrast, the mass of the $\zp$ boson would be $0.45 v_{\chi}$. However, one can assume different values for $f$ parameter, allowing scalars to be lighter than the $\zp$, as has been explored in \cite{Oliveira:2022vjo}. Notice that even if they are indeed lighter than the $\zp$, this does not warrant a larger FCNC effect because the magnitude of the FCNC induced by the scalar fields will be subject to arbitrary choices of the couplings in the scalar potential, see Appendix B of \cite{Cogollo:2012ek}. Thus, in summary, scalar fields usually give rise to relatively meager FCNC effects. Moreover, FCNC arising from scalar fields are necessarily less predictive than the ones stemming from gauge interactions mediated by the $\zp$ field. Albeit, one can in principle overlook all these facts and tune the couplings in the scalar potential in such a way as to enhance the FCNC effects coming from scalar fields and potentially make them the dominant contribution.

Now we have reviewed the key aspects of the model, we will concentrate on the main source of flavor-changing neutral current, namely the $\zp$ gauge boson.

\begin{figure*}[t]
  \centering
    \includegraphics[scale=0.3]{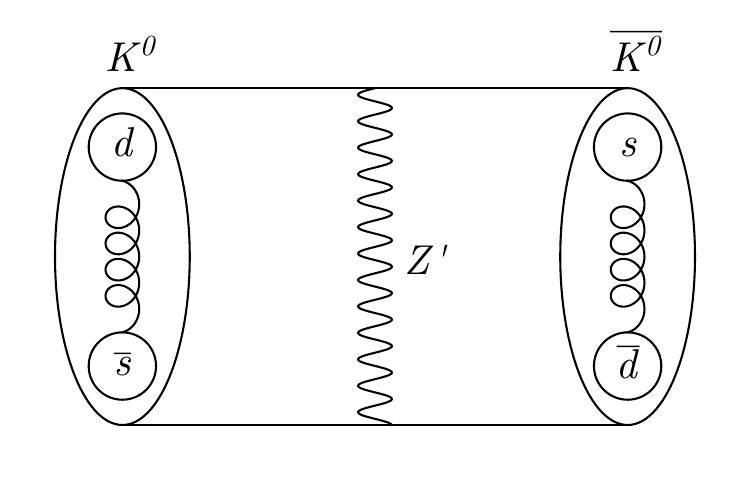}
    \includegraphics[scale=0.3]{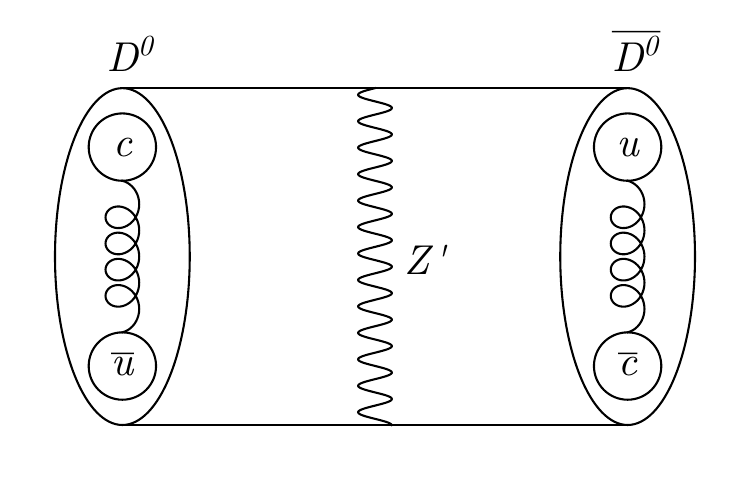}
    \includegraphics[scale=0.3]{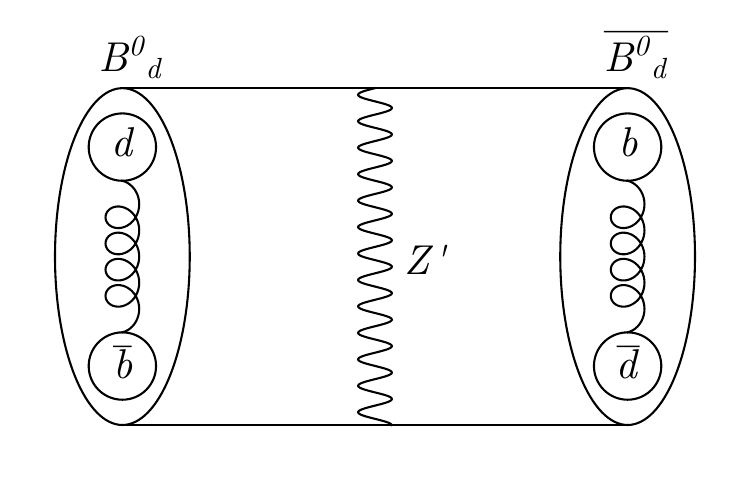}
    \includegraphics[scale=0.3]{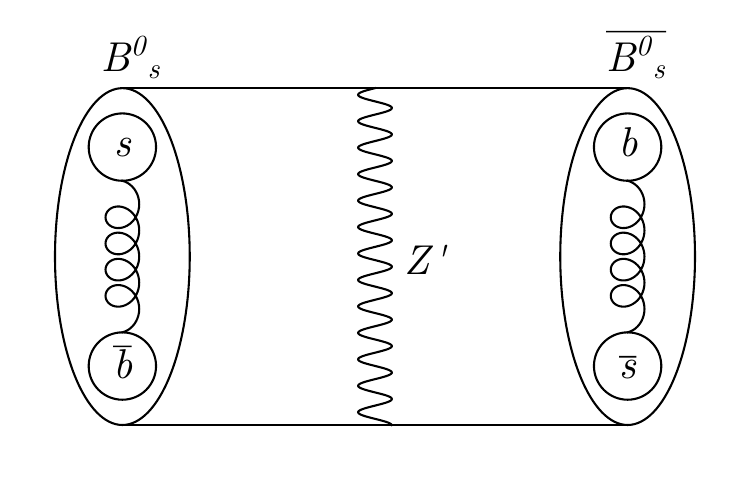}
    \caption{Feynman diagrams illustrating how the $\zp$ gauge boson changes the mass difference of the four meson systems under investigation, namely $K^0-\bar{K}^0$, $D^0-\bar{D}^0$, $B^0_d-\bar{B}^{0_d}$, and $B^0_s-\bar{B}^{0_s}$.}
    \label{fig:mesons}
\end{figure*}

\section{FCNC in the 3-3-1}

Flavor-changing neutral current is a common feature in 3-3-1 models because gauge anomaly cancellation requires one of the fermion generations to transform differently than the others. This requirement naturally induces a flavor-changing neutral current once one rotates the quark flavors and introduces the CKM matrix. In other words, the $\zp$ does not have universal couplings to quarks, and thus flavor changing interactions arise. This is key because the Z boson does not induce flavor changing interactions in the SM, conversely to the charged current mediated by the W boson. Flavor-changing interactions in the SM model occurs through the charged current. Thus, flavor changing interactions induced by a $W^\prime$ would be swamped by numerous W boson interactions. Therefore, it is wise to investigate flavor-changing interactions mediated by a neutral gauge boson, as they are not masked by a large SM effect. We remark that in the 3-3-1 models, we have additional neutral gauge bosons, namely the $W^{\prime^\pm}$,$U^{0}$, and $U^{0\dagger}$. Nevertheless, they do not generate FCNC. Thus, we focus on $\zp$ field.

As we have explained earlier, mesons mass systems are great laboratories to probe such flavor-changing interactions because $\zp$ fields can induce sizeable flavor transitions, impacting the mass difference of meson systems \cite{Misiak:1997ei, Barenboim:2000zz, Datta:2008qn}, see Fig.~\ref{fig:mesons}.  We would like to stress again that
FCNC seeded by scalar fields are typically suppressed compared to those generated by $\zp$ bosons because these scalars are typically heavier than the $\zp$ field, see \cite{Cogollo:2012ek, Mizukoshi:2010ky} and references therein. In fact, it has been shown in \cite{Cogollo:2012ek} that two neutral scalars can induce sizeable FCNC, but their masses go as $m \sim v_{\chi}$, rendering them relatively heavier than the $\zp$. Besides, their contribution to FCNC adds an extra systematic effect to the 3-3-1 prediction, which are the Yukawa couplings and the couplings in the scalar potential. Therefore, there is no predictivity regarding neutral scalar contributions to FCNC. Anyway, this aspect has been explored in \cite{Oliveira:2022vjo}. Lastly, the scalars in the 3-3-1 models do not offer clean collider signals and their couplings to SM fermions are proportional to Yukawa couplings, which result in suppressed production rates at colliders. Consequently, the interplay between FCNC and collider physics is lost. Albeit, in principle, one can certainly fine-tune the couplings in the scalar potential and generate a scalar lighter than the $\zp$ boson making the reasoning in \cite{Oliveira:2022vjo} valid, but thus far, this has not been explicitly proven. For all these reasons, we focus on the $\zp$ field.

In this way, after developing the covariant derivative, we find the following currents,

\begin{eqnarray}
\label{zprimau-1}
\ensuremath{\mathcal{L}}^{\zp}_{u} & = & \frac{g}{2C_{W}}\left(\frac{(3-4S_{W}^{2})}{3\sqrt{3-4S_{W}^{2}}}\right) \left[\bar{u}_{iL}\gamma_{\mu}u_{iL}\right] \zp_{\mu} \nonumber \\
 & -  &  \frac{g}{2C_{W}}\left(\frac{6(1-S_{W}^{2})}{3\sqrt{3-4S_{W}^{2}}}\right)\left[\bar{u}_{3L}\gamma_{\mu}u_{3L}\right] \zp_{\mu},\\
\ensuremath{\mathcal{L}}^{\zp}_{d} & = & \frac{g}{2C_{W}}\left(\frac{(3-4S_{W}^{2})}{3\sqrt{3-4S_{W}^{2}}}\right)
\left[\bar{d}_{iL}\gamma_{\mu}d_{iL}\right] Z_{\mu}^{\prime} \nonumber \\
& - &  \frac{g}{2C_{W}}\left(\frac{6(1-S_{W}^{2})}{3\sqrt{3-4S_{W}^{2}}}\right)
\left[\bar{d}_{3L}\gamma_{\mu}d_{3L}\right] Z_{\mu}^{\prime},
\label{eq:FCNC1} 
\end{eqnarray} 
with $i=1,\, 2,$, indicate the generation indices, and $C_{W} \equiv cos\theta_{W}$, $S_{W} \equiv sin\theta_{W}$, with $\theta_{W}$ being the Weinberg angle. Note that Eqs.~\eqref{zprimau-1} and \eqref{eq:FCNC1} are in the mass eigenstate basis, and once we rotate to the flavor basis, FCNC arises. The mass eigenstate and flavor bases are connected as follows, 
\begin{eqnarray}
\left(\begin{array}{c}
u\\
c\\
t
\end{array}\right)_{L,R}
=
V_{L,R}^{U}
\left(\begin{array}{c}
u^{\prime}\\
c^{\prime}\\
t^{\prime}
\end{array}\right)_{L,R},
\nonumber
\left(\begin{array}{c}
d\\
s\\
b
\end{array}\right)_{L,R}
=
V_{L,R}^{D}
\left(\begin{array}{c}
d^{\prime}\\
s^{\prime}\\
b^{\prime}
\end{array}\right),
\label{misturaquarksMP}
\end{eqnarray} 
where $V_{L,R}^{U}$ and $V_{L,R}^{D}$ are $3\times 3$ unitary matrices, which determine the Cabibbo-Kobayashi-Maskawa (CKM) matrix  $V_{CKM}= V^{U\dagger}_{L} V^{D }_{L}$, 
known to be \cite{Workman:2022ynf}
\begin{eqnarray}
&V&_{CKM} = 
\label{Vckm}\\
&& \left(
\begin{array}{ccc}
0.97435  \scriptstyle{\pm 0.00016} & 0.22500  \scriptstyle{\pm 0.00067} & 0.00369  \scriptstyle{\pm 0.00011} \\
0.22486  \scriptstyle{\pm 0.00067} & 0.97349  \scriptstyle{\pm 0.00016} & 0.04182_{-0.000074}^{+0.00085} \\
0.00857_{-0.00018}^{+0.00020} & 0.04110_{-0.00072}^{+0.00083} & 0.999118_{-0.000036}^{+0.000031}
 \end{array}\right). \nonumber
\end{eqnarray}

After rotation, we get the tree level $\zp$ mediated neutral current interactions,
\begin{eqnarray}
 \mathcal{L}^{K_0-\bar{K}_0}_{\zp \, eff} &=& G^{\prime} \frac{M_Z^2}{M_{\zp}^2}|(V_{L}^{D})_{31}^*(V_{L}^{D})_{32}|^2|\bar{d}^{\prime}_{1L}\gamma_\mu d^{\prime}_{2L}|^2, \nonumber\\
 \mathcal{L}^{D_0-\bar{D}_0}_{\zp \, eff} &=& G^{\prime} \frac{M_Z^2}{M_{\zp}^2}|(V_{L}^{U})_{31}^*(V_{L}^{U})_{32}|^2|\bar{u}^{\prime}_{1L}\gamma_\mu u^{\prime}_{2L}|^2, \nonumber\\
 \mathcal{L}^{B^0_d-\bar{B}^0_d}_{\zp \, eff} &=& G^{\prime} \frac{M_Z^2}{M_{\zp}^2}|(V_{L}^{D})_{31}^*(V_{L}^{D})_{33}|^2|\bar{d}^{\prime}_{1L}\gamma_\mu d^{\prime}_{3L}|^2,\nonumber\\
 \mathcal{L}^{B^0_s-\bar{B}^0_s}_{\zp \, eff} &=& G^{\prime} \frac{M_Z^2}{M_{\zp}^2}|(V_{L}^{D})_{32}^*(V_{L}^{D})_{33}|^2 |\bar{d}^{\prime}_{2L}\gamma_\mu d^{\prime}_{3L}|^2,\nonumber
 \label{eq:FCNC8}
\end{eqnarray}and consequently \cite{GomezDumm:1994tz,Long:1999ij,Benavides:2009cn},
\begin{eqnarray}
 (\Delta m_K)_{\zp} &=& G^{\prime} \frac{M_Z^2}{M_{\zp}^2} |(V_{L}^{D})_{31}^*(V_{L}^{D})_{32}|^2 f_K^2 B_K \eta_K m_K, \nonumber\\
 (\Delta m_D)_{\zp} &=& G^{\prime} \frac{M_Z^2}{M_{\zp}^2}|(V_{L}^{U})_{31}^*(V_{L}^{U})_{32}|^2 f_D^2 B_D \eta_D m_D, \nonumber\\
 (\Delta m_{B_d})_{\zp} &=& G^{\prime} \frac{M_Z^2}{M_{\zp}^2}|(V_{L}^{D})_{31}^*(V_{L}^{D})_{33}|^2 f_{B_d}^2 B_{B_d} \eta_{B_d} m_{B_d}, \nonumber\\ 
 \nonumber
 (\Delta m_{B_s})_{\zp} &=& G^{\prime} \frac{M_Z^2}{M_{\zp}^2}|(V_{L}^{D})_{32}^*(V_{L}^{D})_{33}|^2 f_{B_s}^2 B_{B_s} \eta_{B_s} m_{B_s}.\\
 \label{eq:FCNC3}
\end{eqnarray}where $G^{\prime}= \frac{4 \sqrt{2}G_F C_W^4}{3-4 S_{W}^2}$, with $G_F$ being the Fermi constant, $B_K,B_D,B_B$ the bag parameters, $f_K,f_D,f_B$ the decay constants, and $\eta_K,\eta_D,\eta_B$  the QCD leading order correction obtained in \cite{Misiak:1997ei,Barenboim:2000zz,Datta:2008qn,Dowdall:2019bea,Zyla:2020zbs,Branco:2021vhs,NguyenTuan:2020xls}, and $m_K,m_D,m_B$ the masses of the mesons. In {\it Table} \ref{tableI} we summarize the values of these parameters.   

Our reasoning to constrain new physics contributions to the mass mixing systems goes as follows:
\begin{enumerate}[(i)]
    \item The experimental mass difference of the $K_0-\bar{K}_0$ system is given by $\left(\Delta m_K\right)_{exp}$;
    \item The SM prediction $\left(\Delta m_K\right)_{SM}$  (see \tab{tableI}) has a good agreement with the experimental, but errors are not included in the SM prediction. We find different values for the SM contribution in the literature;
    \item Therefore, instead of imposing  $\left(\Delta m_K\right)_{SM}$ + $\left(\Delta m_K\right)_{\zp} <  \left(\Delta m_K\right)_{exp}$ as done in previous works \cite{Machado:2013jca,Correia:2015tra,CarcamoHernandez:2021osw,CarcamoHernandez:2021yev,Hernandez:2021uxx,Hernandez:2021xet}, we enforce the $\zp$ contribution to be smaller than the statistical error bar {\it Table} \ref{tableI}. In this way, our conclusions are less sensitive to theoretical uncertainties and are driven by experimental measurements.
    \item We follow the same strategy for all four meson systems.
    \item In summary, we impose,     
\begin{eqnarray}
 &&(\Delta m_K)_{\zp}  <  0.006 \times 10^{-12}\, {\rm MeV}, \nonumber\\
 &&(\Delta m_D)_{\zp}   <  2.69 \times 10^{-12}\,  {\rm MeV}, \nonumber\\
 &&(\Delta m_{B_d})_{\zp}  < 0.013 \times 10^{-10} \, {\rm MeV}, \nonumber\\
 &&(\Delta m_{B_s})_{\zp}  < 0.0013 \times 10^{-8}\,  {\rm MeV}.
 \label{eqmesonexp}
\end{eqnarray}
     
\end{enumerate}

We remind the reader that that $\zp$ boson mediates FCNC at tree-level through Eq.~\eqref{eq:FCNC3} and for this reason, we will be able to severely constrain the mass of this particle. An advantage of working in the scope of a 3-3-1 model is that $\zp$ boson couples to SM fields proportional to the $SU(2)_L$ gauge coupling. The only unknown quantities are the mixing matrices and the $\zp$ mass. 

\begin{table}[!h]
{\color{blue} Input parameters}\\
\begin{tabular}{|c|}
\hline
$\left(\Delta m_K\right)_{exp}=\left(3.484\pm 0.006\right)\times 10^{-12}$ MeV \\
$\left(\Delta m_K\right)_{SM}=3.483\times 10^{-12}$ MeV \\
 $m_K=\left(497.611\pm 0.013\right)$ MeV  \\
 $\sqrt{B_K}f_K=131$ MeV \\
 $\eta_K=0.57$ \\
\hline

\hline
$\left(\Delta m_D\right)_{exp}=\left(6.25316_{-2.8962}^{+2.69873}\right)\times 10^{-12}$~MeV  \\
$\left(\Delta m_D\right)_{SM}=10^{-14}$~MeV  \\
$m_D=\left(1865\pm 0.005\right)$ MeV  \\
$\sqrt{B_D}f_D=187$ MeV  \\
$\eta_D=0.57$  \\
\hline

\hline
$\left(\Delta m_{B_d}\right)_{exp} =\left(3.334\pm 0.013\right) \times 10^{-10}$ MeV \\ 
$\left(\Delta m_{B_d}\right)_{SM} =\left(3.653\pm 0.037\pm 0.019\right) \times 10^{-10}$ MeV \\
 $m_{B_d}=\left(5279.65\pm 0.12\right)$ MeV \\
$\sqrt{B_{Bd}}f_{Bd}=210.6$ MeV \\
$\eta_{B_d}=0.55$ \\
\hline

\hline
$\left(\Delta m_{B_s}\right)_{exp} =\left(1.1683\pm 0.0013\right) \times 10^{-8}$ MeV \\ 
$\left(\Delta m_{B_s}\right)_{SM} =\left(1.1577\pm 0.022\pm 0.051\right) \times 10^{-8}$ MeV \\
 $m_{B_s}=\left(5366.9\pm 0.12\right)$ MeV \\
$\sqrt{B_{B_s}}f_{B_s}=256.1$ MeV  \\
$\eta_{B_s}=0.55$ \\
\hline
\end{tabular}
\caption{Meson masses \cite{CPLEAR:1998zfe,Artuso:2015swg,Jubb:2016mvq,Wang:2018csg,Lenz:2019lvd,HFLAV:2019otj,Aoki:2021kgd} and the values of the bag parameters \cite{Workman:2022ynf,Aoki:2021kgd}.}
\label{tableI}
\end{table}

We will assume two different parametrizations of the mixing matrices that yield significant changes in the new physics contribution to the mass difference systems. In this way, we can assess the impact that such parametrizations. We adopt  {\it parametrization 1},

\begin{equation}
\label{para1}
V_{L}^{D}=V_{R}^{D}=\left(
\begin{array}{ccc}
  0.972 & 0.5 &  0.46 \\
  0.45 & 1.00 & 0.88 \\
  0.1 & 0.1 &  1.01
 \end{array}\right)
\end{equation}
and,
$$
V_{L}^{U}=V_{R}^{U}=\left(
\begin{array}{ccc}
1.18622007& -0.22070355& -0.09032872\\
-0.34446205&  1.17174168& -0.01837301\\
-0.23647983& -0.87899906&  1.04637372
 \end{array}\right),
$$        
and {\it parametrization 2},
\begin{equation}
\label{para2}
V_{L}^{D}=V_{R}^{D}=\left(
\begin{array}{ccc}
  0.972 & 0.5 &  0.46 \\
  0.45 & 1.00 & 0.88 \\
  0.1 & 0.0001 &  1.01
 \end{array}\right)
\end{equation}
and,
$$
V_{L}^{U}=V_{R}^{U}=\left(
\begin{array}{ccc}
 1.19772759& -0.17792992& -0.1412471 \\
0.37384218&  1.06253529&  0.11162792\\
-0.21612235& -0.80332999&  0.95629613
 \end{array}\right).
$$ 

Knowing the entries of the up-quark and down-quark mixing matrices $V^u_L$ and $V^d_L$, we determine the $\zp$ contribution to the mass difference of the meson systems and consequently place a lower mass bound. We adopt these parametrizations because they yield very strong and very conservative 3-3-1 contributions to FCNC processes, respectively, while keeping the CKM matrix in agreement with the data. With this information at hand, we use Eq.\eqref{eq:FCNC3} combined with Eq.\eqref{eqmesonexp} to plot our findings in Figs.~\eqref{finalplot1}-\eqref{finalplot4}.  Using these parametrizations, one can assess the systematic uncertainty embedded in FCNC studies. In other words, FCNC alone is not robust enough.

Before discussing our results, it is important to put them into context with current and future collider bounds. To do so, we address those limits below. 

\section{Dilepton Resonance Searches at the LHC}

$\zp$ gauge bosons are often targets of experimental searches going from low to the multi-TeV mass range \cite{Alves:2015mua, Alves:2015pea, Klasen:2016qux, Kahlhoefer:2019vhz}. In the TeV range, which is the focus of our study, $\zp$ gauge bosons that feature sizeable couplings to fermions can leave a clear signature at the LHC in the form of di-jet and dilepton events. In the 3-3-1 model, the $\zp$ has similar couplings to quarks and leptons. As dilepton events have relatively good signal efficiencies and acceptance and a well-controlled background originating primarily from Drell-Yann processes \cite{Jezo:2014wra, Jezo:2015rha, Klasen:2016qux}, tighter constraints on the $\zp$ mass are found compared to di-jet events. There have been experimental searches for $\zp$ gauge bosons belonging to the 3-3-1 symmetry in the past \cite{Coutinho:2013lta, Salazar:2015gxa}. The most recent analysis taking advantage of the full dataset from LHC was carried out in \cite{Alves:2022hcp}. We consider the most conservative bounds, which is the third benchmark scenario presented in {\it Table IV} of \cite{Alves:2022hcp}. The LHC bound was based on an integrated luminosity of $\mathcal{L}=139 fb^{-1}$ with $\sqrt{s}=13$~TeV, whereas for the High-Luminosity LHC setup $\mathcal{L}=3000 fb^{-1}$ with $\sqrt{s}=14$~TeV.  For the High-Energy LHC the latter luminosity was adopted, but using $\sqrt{s}=27$~TeV. In summary, we used,

\begin{itemize}
\item $M_{\zp} \geq 4$~TeV, LHC 13~TeV
\item $M_{\zp} \geq 5.6 $~TeV, HL-LHC 14~TeV
\item $M_{\zp} \geq 9.6 $~TeV, HE-LHC 27~TeV
\item $M_{\zp} \geq 27$~TeV, FCC-hh 100TeV \normalcolor
\end{itemize} 

\begin{figure*}[!ht]
 \centering
 \includegraphics[scale=0.6]{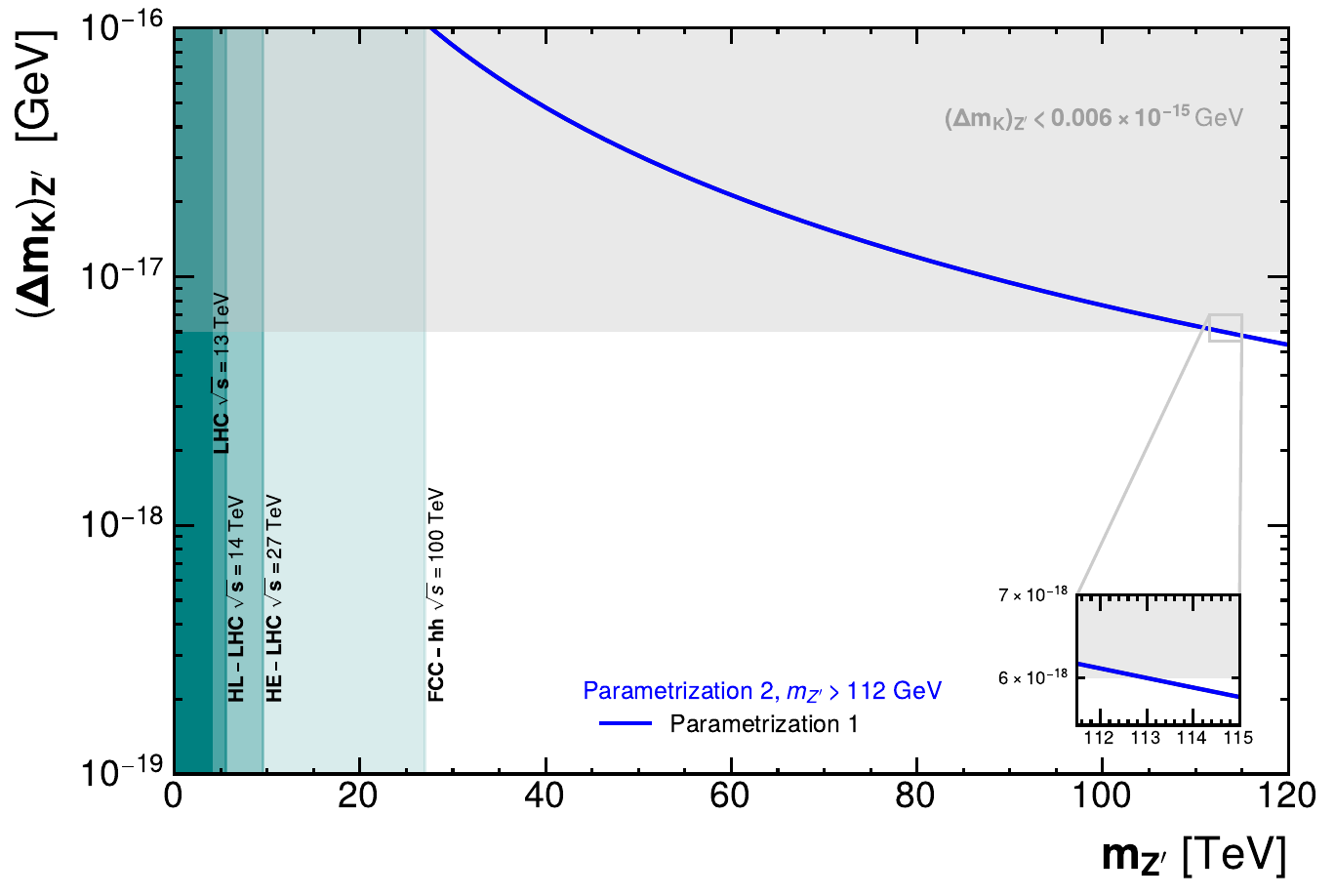}
 \caption{The $\zp$ contribution to the $(\Delta m_K)_{\zp}$ as a function of it mass (see \eq{eq:FCNC3}) for the {\it parametrization 1}, \eq{para1} (blue solid curve), and {\it parametrization 2}, \eq{para2}. The silver region corresponds to the FCNC exclusion region. We overlaid current and projected colliders bounds. Note that parameterization 2 is not shown in the plot because the lower bound of the $\zp$ boson mass in both parameterizations is substantially different. See text for details.}
 \label{finalplot1}
\end{figure*}

\begin{figure*}[!]
 \centering
 \includegraphics[scale=0.6]{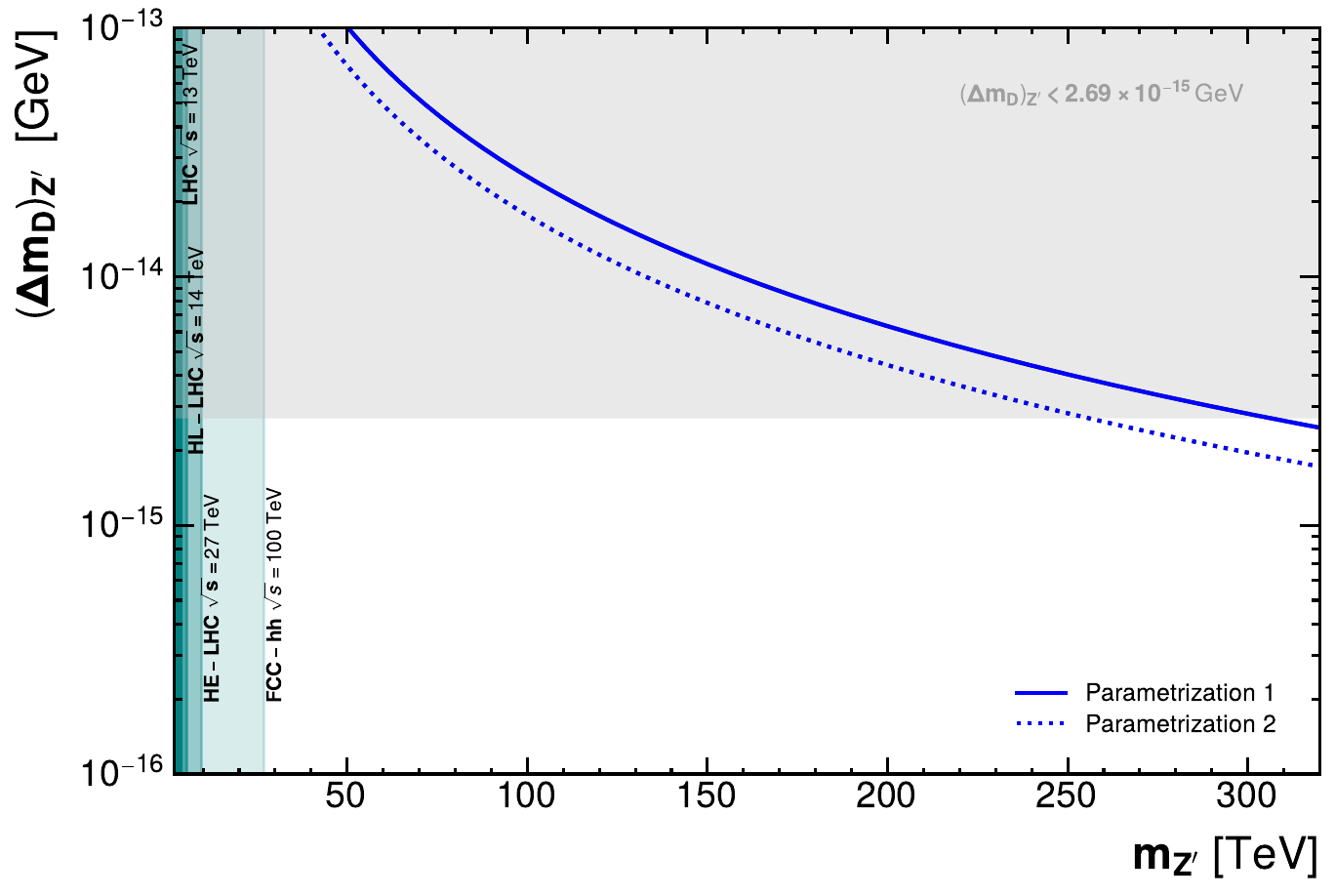}
 \caption{The blue curves correspond to $\zp$ contribution to the $(\Delta m_D)_{\zp}$ as a function of it mass (see \eq{eq:FCNC3}), for the {\it parametrization 1}, \eq{para1}, and {\it parametrization 2}, \eq{para2}. The silver region corresponds to the FCNC exclusion region. The lower mass bound for the {\it parametrization 2} is $m_{\zp} >256$~TeV. We overlaid current and projected colliders' bounds. See text for details.}
  \label{finalplot2}
\end{figure*}

\begin{figure*}[!]
 \centering
 \includegraphics[scale=0.6]{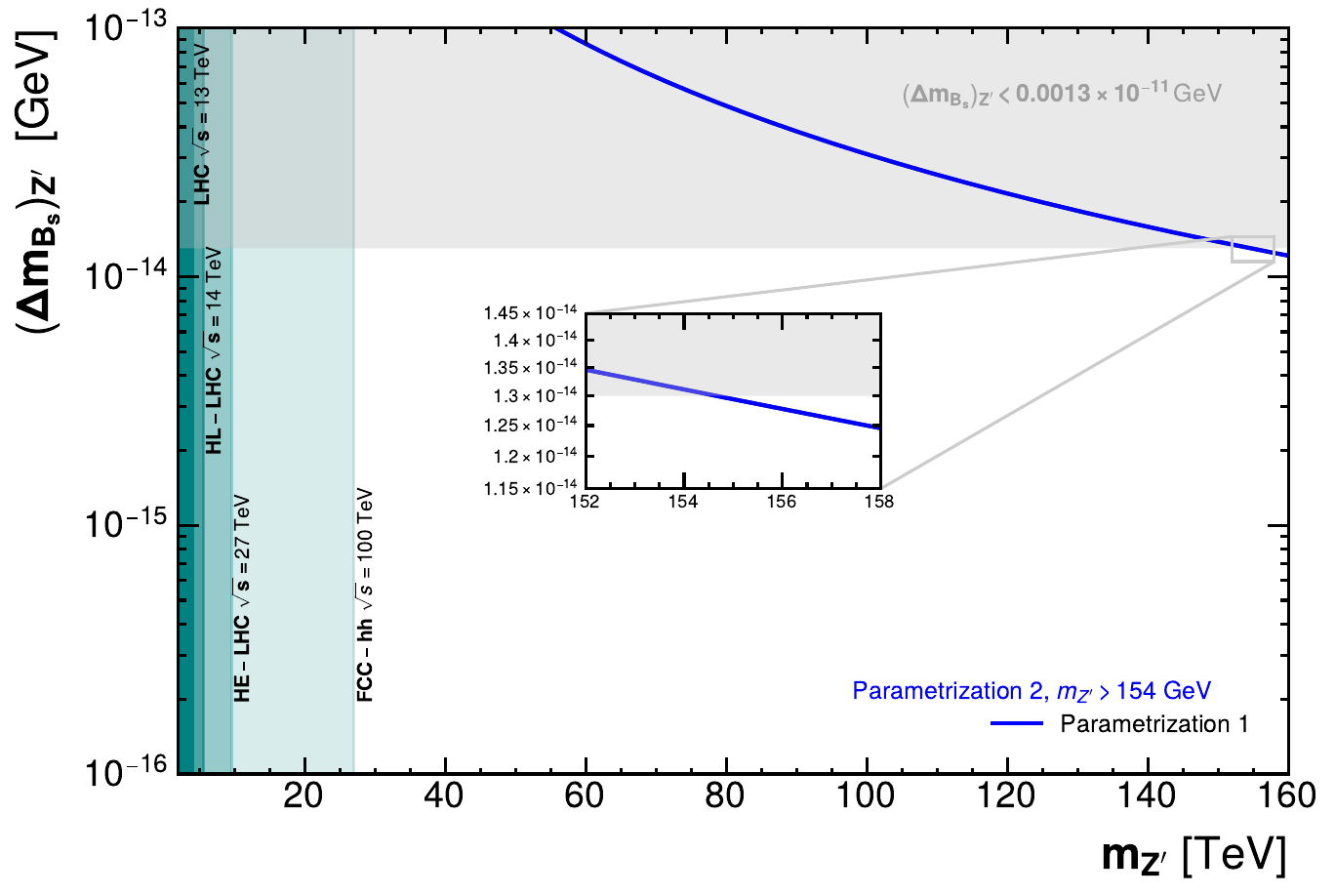}
 \caption{The solid blue line corresponds to $\zp$ contribution to the $(\Delta m_{B_s})_{\zp}$ as a function of it mass (see \eq{eq:FCNC3}), for the {\it parametrization 1}, \eq{para1}, and {\it parametrization 2}, \eq{para2}. The silver region corresponds to the FCNC exclusion region. The lower mass bound for the {\it parametrization 1} is $m_{\zp} >154$~TeV.
 We overlaid current and projected colliders bounds. See text for details.}
  \label{finalplot3}
\end{figure*}

\begin{figure*}[!]
 \centering
 \includegraphics[scale=0.6]{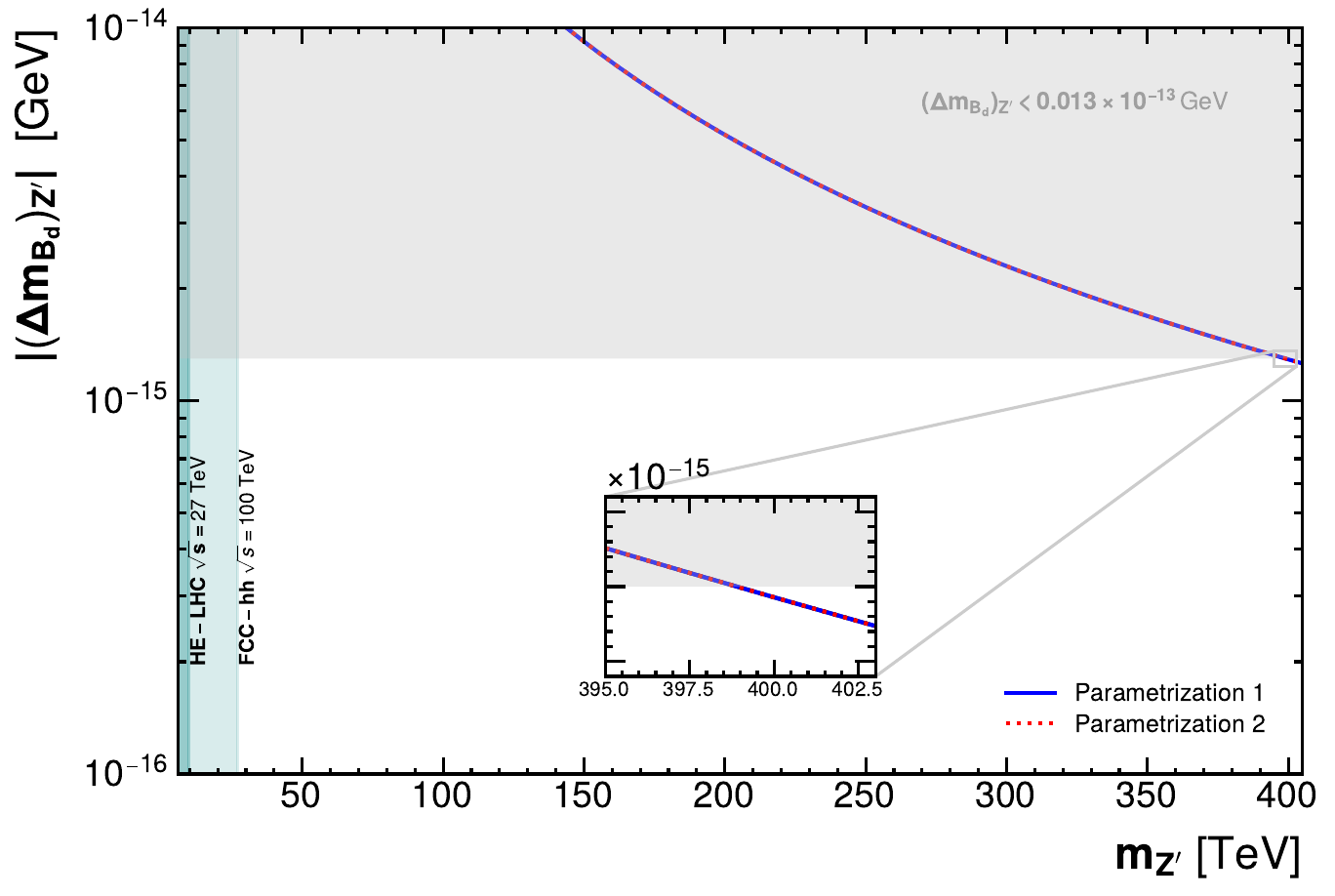}
 \caption{The solid blue and dotted red lines  correspond to $\zp$ contribution to the $(\Delta m_{B_d})_{\zp}$ as a function of it mass (see \eq{eq:FCNC3}), for the two parametrizations of the $V_{L}^{D}$ matrix, see \cref{para1,para2}. The silver region corresponds to the FCNC exclusion region. We overlaid current and projected colliders bounds. See text for details}
  \label{finalplot4}
\end{figure*}

These limits are exhibited in Figs.\eqref{finalplot1}-\eqref{finalplot4}. We have gathered enough information to discuss our results.

\section{Discussion}
  
For the $K^0 - \bar{K}^0$ system, the results are summarized in Fig.~\ref{finalplot1}. The silver region corresponds to the region in which the $\zp$ contribution exceeds the experimental error (see Eq.~\eqref{eqmesonexp}). One can see that the parametrizations one and two give rise to distinct bounds on the $\zp$ mass. Adopting {\it parametrization 1} we find $m_{\zp} >  113$~TeV, whereas using {\it parametrization 2} we get $m_{\zp} > 112$~GeV. We superimposed the LHC 13TeV bound as well as projections for the HL-LHC, HE-LHC and FCC-hh collider.

Regarding $D^0 - \bar{D}^0$ system, Fig.~\ref{finalplot2}, we get $m_{\zp} > 307$~TeV for {\it parametrization 1}, and  {\it parametrization 2} we get $m_{\zp} > 256$~TeV. We superimposed the LHC 13TeV bound as well as projections for the HL-LHC, HE-LHC, and FCC-hh collider.
As for the $B_{s}^{0} - \bar{B_{s}}^0$ system, Fig.~\ref{finalplot3}, we obtain $m_{\zp} > 154$~TeV for {\it parametrization 1}, and  {\it parametrization 2} we get $m_{\zp} > 154$~GeV. 
Lastly, for the $B_{d}^{0} - \bar{B_{d}}^0$ system, Fig.~\ref{finalplot4}, we find $m_{\zp} > 400$~TeV for both parametrizations.

We highlight that in Figs.~\ref{finalplot1} and \ref{finalplot3} the lower bound on the $\zp$ boson mass rising from parametrization 2, is too weak, falling out of the plot range. Thus, it does not appear in the figures. It is clear from our findings that one ought to consider all four meson systems at the same time because one can randomly pick a parametrization designed to suppress the new physics contribution for a given meson system. Without a general approach over FCNC no solid conclusions can be drawn. Moreover, for the parametrization explored in this work, the $B_{d}^{0} - \bar{B_{d}}^0$ system is the most constraining. We remind the reader that our lower mass bounds are driven by experimental errors, as discussed in Eq.~\eqref{eqmesonexp}. It is exciting to see the interplay between future colliders and FCNC because depending on the parametrization used, FCNC can offer a most restrictive probe than future colliders.

We highlight that our conclusions are also applicable to the 3-3-1 model with heavy neutral leptons because the neutral current is identical \cite{Mizukoshi:2010ky,Ruiz-Alvarez:2012nvg, Profumo:2013sca,Kelso:2013nwa}. One should have in mind,  depending on the parametrization adopted, FCNC does lead to a lower mass bound much stronger than the LHC and even future colliders. Hence, one cannot overlook the $\zp$ contributions to FCNC processes.

Having in mind the complementary aspect between flavor physics and colliders, we discuss recent flavor anomalies in the context of 3-3-1 models.

\section{ Flavor anomalies}

\subsection{$b\rightarrow s$ transitions }
The $b\rightarrow s$ transitions not consistent with the SM predictions have been observed in the LHCb data \cite{LHCb:2014vgu,LHCb:2015svh,Descotes-Genon:2013wba,Hiller:2014yaa,Descotes-Genon:2015uva}, which has triggered a multitude of new physics studies in the context of $\zp$ models. Some of them which are of interest to us reside on the $SU(3)_C \times SU(3)_L \times U(1)_N$ gauge group. It is true that there are several ways to arrange the fermion content under this gauge symmetry, and these arrangements have an impact on the precise neutral current mediated by $\zp$ boson. However, the impact is minimal as far as collider physics goes. If there are new exotic fermions that couple to the $\zp$ boson and are sufficiently light, the collider limits based on dilepton searches will be weakened due to the presence of a new and significant decay mode. Besides collider physics, the mass difference of the four meson systems also places a bound on the $\zp$ mass. That said, we will assess whether these interpretations to explain the $b \rightarrow s$ anomaly are indeed viable. In \cite{Descotes-Genon:2017ptp}, the authors considered a model similar to ours but with five lepton generations. The SM quarks possess the same quantum numbers as ours. If the exotic leptons in \cite{Descotes-Genon:2017ptp} are sufficiently heavy to not contribute to the $\zp$ decay width, the aforementioned collider limits are also applicable. In order to fit  the $b\rightarrow s \ell \ell$ anomaly, according to the recent global fits one needs $C_9^\mu=-C_{10}^\mu \simeq -0.6$. Being $C_9^\mu$ and $C_{10}^\mu$ the Wilson coefficients that contribute to new physics present in the effective Hamiltonian described in Eq. (30) of \cite{Descotes-Genon:2017ptp}. In \cite{Descotes-Genon:2017ptp} however, two quantities are important  $r_{B_s}$ and $C_9^\mu$, with the former controlling the bound from the $B_s$ mixing and the latter the $b\rightarrow s \ell \ell$ anomaly. The 3-3-1 model could explain the LHCb anomaly without being excluded by $B_s$ mixing if $C_9^\mu=-C_{10}^\mu \simeq -0.6$ and $r_{B_s}\simeq 0.1$. However, $r_{B_s}=347 \times 10^3 (m_W/m_{\zp})^2 d^2$, and $C_9^\mu =11.3 \times 10^3 (m_W/m_{\zp})^2 d$, where $d=-0.005$ is a parameter that depends on the entries of the quark mixing matrices relevant for $B_s$ mixing ($d = (V_{L}^{D})_{32}^*(V_{L}^{D})_{33}$). This value was assigned to obey the current bound. When we use our parameterizations, the parameter d takes the following values: $0.101$ and $0.000101$ for the parameterizations 1 and 2 respectively.

Given the current LHC bound on the $\zp$ mass, $\sim 4$~TeV, one cannot explain simultaneously address the LHCb anomaly and respect the LHC lower mass bound. We emphasize that this $4$~TeV bound relies on the assumption that there are no extra decay modes besides the usual 3-3-1 field content. Hence, a way to circumvent our conclusion is allowing the extra leptons added in \cite{Descotes-Genon:2017ptp} to be sufficiently light to decrease the $\zp$ branching ratio into charged leptons and consequently weaken the LHC bound. This is a non-trival task knowing that these leptons are chiral leptons, thus can be produced via SM gauge bosons at colliders, and consequently are subject to strong collider bounds \cite{Gogoladze:2010in,Altmannshofer:2013zba,Ma:2014zda,Cogollo:2015fpa,Gopalakrishna:2015wwa,Helo:2018qej,Rappoccio:2018qxp,Guedes:2021oqx}.

In \cite{Addazi:2022frt}, the authors investigated a similar 3-3-1 model and advocated that existing collider bounds on the $\zp$ gauge boson belonging to 3-3-1 model could be significantly lowered if all $\zp$ decay channels modes are included. The possible 3-3-1 decay channels have already been included in \cite{Alves:2022hcp}. Once more, a weakening of the LHC bound would require the chiral leptons introduced in \cite{Addazi:2022frt} to be sufficiently light. Our reason to disfavor this possibility was mentioned above. 

\subsection{$R(D^{\ast})$ Anomaly}    

 In \cite{Wei:2017ago}, the authors considered an exotic field content based on the 3-3-1 symmetry, and focused on the charged Higgs contribution to the semileptonic B-meson decay, particularly on $R(D^{\ast})$ the anomaly reported by BABAR, Belle and LHCb. However, the authors argue that they can take the charged Higgs mass below 1~TeV while keeping the gauge boson masses at sufficiently high scales. It has been shown that despite being a scalar, its mass is naturally predicted to be around the energy scale at which the 3-3-1 symmetry is spontaneously broken, unless ones invoke a fine-tuning in the quartic scalar couplings 
 \cite{Oliveira:2022vjo}. In other words, the mass of the charged scalar is around $v_{\chi}$. Therefore, given the collider bounds, and the FCNC bounds we derived, the proposed 3-3-1 explanation to the $R(D^{\ast})$ anomaly is disfavored.

\section{Conclusions}

We have studied FCNC in a 3-3-1 model using the four meson systems, namely $K^{0}-\bar{K}^{0}$, $D^{0}-\bar{D}^{0}$, $B^0_d-\bar{B^0}_d$ and $B^0_s-\bar{B^0}_s$. We derived lower mass bounds that range from 112 GeV up to 400 TeV using different parametrizations of the quark mixing matrices to solidly show that constraints stemming from FCNC are subject to large systematic uncertainties. We have shown that a robust assessment of FCNC should consider the four meson systems because specific parametrizations of the quark mixing matrices can suppress new physics effects at one of the meson systems. However, as the CKM matrix should be preserved, these parametrizations tend to enhance FCNC effects on the other mesons.  We carried out a study based on the $\zp$ contributions to FCNC, as the scalars are typically much heavier than the $\zp$ field, their corrections to FCNC are  subdominant. Considering only gauge interactions the systematic effects already drive the new physics sensitivity, let alone the scalar fields whose contributions depend on arbitrary choices of the Yukawa couplings and scalar potential parameters. In summary, a broader view of FCNC is needed before drawing conclusions. Lastly, we argued that recent anomalies in $b\rightarrow s$ and $R(D^{\ast})$ transitions are disfavored in light of recent collider bounds.

\acknowledgments
We thank Diego Cogollo and Carlos Pires for discussions. This work was financially supported by Simons Foundation (Award Number:884966, AF), FAPESP grant 2021/01089-1, ICTP-SAIFR FAPESP grant 2021/14335-0, Coordenação de Aperfeiçoamento de Pessoal de Nível Superior - Brasil (CAPES) - Finance Code 001, CNPq grant 408295/2021-0, Serrapilheira Foundation (grant number Serra-1912–31613), ANID-Chile FONDECYT 1210378 and 1190845, ANID PIA/APOYO AFB180002 and ANID- Programa Milenio - code ICN2019 044. CS is supported by grant 2020/00320-9, S\~ao Paulo Research Foundation (FAPESP).

\def\bibsection{\section*{References}}
\bibliographystyle{apsrev4-1}
\bibliography{darkmatter}

\end{document}